\input epsf.tex    

\documentstyle[12pt]{article}

\newcommand{\bmat}{\left(\begin{array}}
\newcommand{\emat}{\end{array}\right)}

\def\yzero{\smash{\hbox{$y\kern-4pt\raise1pt\hbox{${}^\circ$}$}}}

\def\beq{\begin{equation}}
\def\eeq{\end{equation}}
\def\beqa{\begin{eqnarray}}
\def\eeqa{\end{eqnarray}}

\def\-{\hphantom{-}}

\def\s2{\frac{1}{2}}

\def\beq{\begin{equation}}
\def\eeq{\end{equation}}
\def\beqa{\begin{eqnarray}}
\def\eeqa{\end{eqnarray}}

\def\ch{{\rm ch \,}}

\def\IF{\relax{\rm I\kern-.18em F}}
\def\II{\relax{\rm I\kern-.18em I}}

\def\cp{{\cal P}}
\def\IC{\bf C}
\def\IZ{\bf Z}
\def\IR{\bf R}
\def\IS{\bf S}
\def\IP{\bf P}
\def\z2z2{$\IC^3/(\IZ_2\times\IZ_2)$}

\def\Dsl{\,\raise.15ex\hbox{/}\mkern-13.5mu D} 

\def\IT{\bf T}

 \def\cp#1{\relax\ifmmode {\IP\kern-2pt{}_{#1}}\else $\IP\kern-2pt{}_{#1}$\=fi}
\newcommand{\drawsquare}[2]{\hbox{%
\rule{#2pt}{#1pt}\hskip-#2pt
\rule{#1pt}{#2pt}\hskip-#1pt
\rule[#1pt]{#1pt}{#2pt}}\rule[#1pt]{#2pt}{#2pt}\hskip-#2pt
\rule{#2pt}{#1pt}}

\newcommand{\fund}{\raisebox{-.5pt}{\drawsquare{6.5}{0.4}}}
\newcommand{\antifund}{\overline{\fund}}

\begin{document}

\makeatletter \@addtoreset{equation}{section} \makeatother
\renewcommand{\theequation}{\thesection.\arabic{equation}}
\pagestyle{empty}
\vspace*{.5in}
\rightline{FT-UAM-02-04}
\rightline{IFT-UAM-CSIC-02-02}
\rightline{NSF-ITP-02-10}
\rightline{\tt hep-th/0201221}
\vspace{2cm}

\begin{center}
\LARGE{\bf D-branes, fluxes and chirality \\[10mm]}
\medskip
\large{Angel M. Uranga \footnote{\tt Angel.Uranga@uam.es} \\[2mm]}
Dpto. de F\'{\i}sica Te\'orica C-XI and 
Instituto de F\'{\i}sica Te\'orica C-XVI \\
Universidad Aut\'onoma de Madrid, 28049 Madrid, Spain \\ [3mm]

\smallskip

\small{\bf Abstract} \\[3mm]
\end{center}

\begin{center}
\begin{minipage}[h]{14.5cm}
{\small 
We describe a topological effect on configurations of D-branes in the
presence of NS-NS and RR field strength fluxes. The fluxes induce the
appearance of chiral anomalies on lower dimensional submanifolds of the
D-brane worldvolume. This anomaly is not associated to a dynamical chiral
fermion degree of freedom, but rather should be regarded as an explicit
flux-induced anomalous term (Wess-Zumino term) in the action. The anomaly
is cancelled by an inflow mechanism, which exploits the fact that fluxes
can act as sources of RR fields. We discuss several applications of this
flux-induced anomaly; among others, its role in understanding anomaly
cancellation in compactifications with D-branes and fluxes, and the
possibility of phase transitions where a chiral fermion disappears from
the D-brane world-volume spectrum, being replaced by an explicit
Wess-Zumino term. We comment on the relation among different mechanisms to 
obtain four-dimensional chirality in string theory.}

\end{minipage}
\end{center}
\newpage
\setcounter{page}{1} \pagestyle{plain}
\renewcommand{\thefootnote}{\arabic{footnote}}
\setcounter{footnote}{0}

\section{Introduction}
\label{intro}

Compactifications of type II string theory / M-theory with field strength
fluxes turned on (see e.g.
\cite{ps,bb,michelson,gvw,drs,tv,gss,cklt,gkp,kst,fp})  are an interesting
class of constructions which may shed new light on several questions of
phenomenological relevance. For instance, the observation that fluxes lead
to warped internal metric suggests the models may be used to generate
exponential hierarchies \cite{drs,gss,gkp}, following the proposal in
\cite{rs}, yielding a possible solution for the hierarchy of scales in
particle physics. On the other hand, the observation that the presence of
fluxes induces tree-level potential terms for diverse moduli
\cite{gvw,tv}, including even the dilaton, provides a possible canonical
mechanism to stabilize them \cite{drs,gkp,kst}.

Several properties of type II compactifications with fluxes have been
studied recently. In particular, the consistency conditions for the
configurations, the amount of supersymmetry preserved, and also the
effective action for the closed string sector fields (e.g. potential for
moduli). However the physics of D-branes, namely of open string sectors, 
in compactifications with fluxes has been much less analyzed. There are
several indications that such physics is nevertheless extremely
interesting; in fact D-branes in the presence of fluxes present
interesting phenomena, for instance the dielectric effect
\cite{dielectric}.
Their systematic analysis is however difficult given the incomplete
knowledge of non-abelian D-brane actions, and the difficulties to obtain
worldsheet results in situations with RR fluxes.

In this paper we center on a particular effect whose analysis is
relatively simple, since it involves only topological couplings. Yet the
consequences of the effect are extremely interesting. In Section
\ref{chicha}, we
show that, in the presence of suitable NS-NS and RR field strength fluxes,
D-branes develop a chiral anomaly localized at lower-dimensional
submanifolds of their volume. This is derived by first showing there
exists an anomaly inflow from the D-brane world-volume towards such
lower-dimensional slices, which signal the existence of an anomaly source.
The source of the anomaly is not, contrary to other familiar situations, 
a chiral fermion degree of freedom, but rather an explicit anomalous 
interaction developed by the D-brane in the presence of the fluxes.

In Section \ref{applic} we present two applications of this effect. It
plays a key role in understanding anomaly cancellation on the world-volume
of D-brane probes in certain configurations with fluxes, and of
cancellation of anomalies in chiral compactifications with fluxes. In
Section \ref{braneflux} we discuss brane/flux transitions and use them to
generate transitions on D-brane world-volumes, in which a dynamical chiral
fermion disappears from the spectrum, leaving behind an explicit 
Wess-Zumino term. Section \ref{global} discusses how to generate global gauge
anomalies using fluxes. Section \ref{conclu} concludes with a discussion
of the interrelation among different mechanisms to generate
four-dimensional chirality in string theory.

\section{Chirality from fluxes}
\label{chicha}

In this section we discuss the effect of flux-induced anomaly on D-branes
in the presence of fluxes. We start with a review of the anomaly inflow
mechanism in Section \ref{inflow}. In Section \ref{chiral} we show that in
configurations of D-branes and NS-NS and RR fluxes, there exists an inflow 
of anomaly from the D-brane worldvolume towards lower dimensional slices 
of its volume. The inflow signals the existence of an anomaly source, 
which we identify as an explicit anomalous interaction, rather than a
dynamical fermion degree of freedom.

\subsection{The anomaly inflow mechanism}
\label{inflow}

Let us give a simplified review of the anomaly inflow mechanism for 
D-branes, as discussed in \cite{inflow}. Consider a D-brane, which 
couples to the RR fields through the action
\beqa
S_{Dp} \, =\, \int_{Dp} \, {\cal C} \,\wedge \, Y(F,R)
\label{CS}
\eeqa
where ${\cal C}$ is a formal sum of RR forms of different degrees, and $Y$ 
is a closed gauge invariant form depending on the worldvolume gauge field 
strength and curvature, in fact
\beqa
Y \, =\, \ch(F)\, \hat{A}(R)^{1/2}
\eeqa
where $\ch$ is the Chern character and $\hat{A}$ is the A-roof genus. 
In what follows, wedge products will be implicit.

We will be interested in situations where the RR fields have a source 
(different form the D-brane itself), so that the field strength ${\cal G}$ 
obeys
\beqa
d{\cal G}\, =\, {\cal Z}
\eeqa
where ${\cal Z}$ is a source form (in fact a formal sum of them), which 
is often a bump form, a delta-function localized on the core of the 
source (typically, but not necessarily, a D-brane).

In this situation, ${\cal C}$ is not well defined, so the D-brane coupling 
(\ref{CS}) is better defined by
\beqa
S_{Dp} \, =\, \int_{Dp}\, {\cal G} \, \,Y^{(0)}(F,R)
\label{CS2}
\eeqa
where we have used the Wess-Zumino descent relation notation, i.e. for 
a closed gauge invariant form $Y$, we define $Y= dY^{(0)}$, $\delta 
Y^{(0)}= dY^{(1)}$, where $\delta$ represents a gauge variation.

In this situation, the action (\ref{CS2}) is not gauge invariant, its 
gauge variation is given by
\beqa
\delta S_{Dp} \, =\, \int_{Dp}\, {\cal G} \, \, \delta 
Y^{(0)}(F,R) 
\, = \, \int_{Dp}\, d{\cal G} \, \, Y^{(1)}(F,R) \, = \, \int_{Dp} 
\, {\cal Z} \, \, Y^{(1)}(F,R)
\eeqa

Recalling the familiar interpretation of $Y^{(1)}$ as the anomaly 
descending from an anomaly polynomial $Y(F,R)$, it is useful to interpret 
this last expression by saying that there exists an inflow of
chiral anomaly from 
the D$p$-brane wolume towards the core of the source (or rather towards 
its intersection with the D$p$-brane volume), with density given by the 
source form ${\cal Z}$. That is the D$p$-brane action is no longer gauge 
invariant, and its gauge variation is localized on the core of ${\cal Z}$.
If ${\cal Z}$ is not bumpy the inflow at each lower dimensional slice is 
proportional to the magnitude of ${\cal Z}$.

This lack of gauge invariance must be compensated by a counteracting 
effect, arising at the core of the source $Z$. As we discuss below, the 
usual situation is that a dynamical chiral fermion arises, whose anomaly 
is cancelled by this inflow. In the following section we argue there exist 
situations where no such dynamical fermion arises, but instead there
exists an explict Wess-Zumino interaction on the D$p$-brane world-volume.

\medskip

\begin{figure}
\begin{center}
\centering
\epsfysize=3.5cm
\leavevmode
\epsfbox{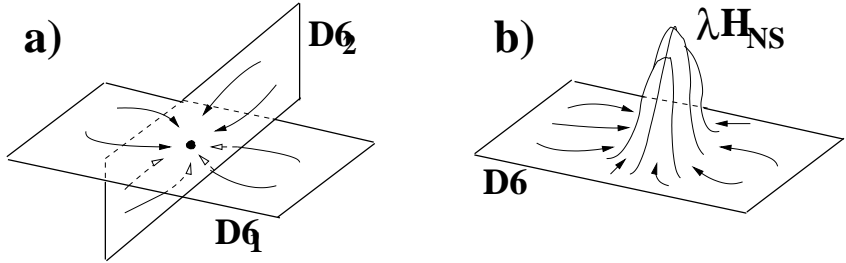}
\end{center}
\caption[]{\small Anomaly inflow for a) D6-branes intersection over a 
four-dimensional submanifold of their world-volume; b) D6-brane in the 
presence of longitudinal $\lambda H_{NS}$ flux.}
\label{fig1}
\end{figure}

A familiar situation where the anomaly inflow is cancelled by the 
appearance of a dynamical chiral fermion degree of freedom is intersecting 
D-branes, see \cite{bdl}. For concreteness consider two stacks of 
D6-branes intersecting over a four-dimensional subspace of their 
world-volumes, Fig \ref{fig1}a. By a slight generalization of the above 
argument (see \cite{inflow}), the world-volume action of each D6-brane 
stack is not 
gauge invariant due to the presence of the other D6-brane stack, which 
acts as a source. This results in an overall gauge variation
\beqa
\delta S_{D6_1+D6_2}\, = \, \int \, [\,Y(F_1,R) \,Y(F_2,R)\,]^{(1)} 
 \, \delta_I
\label{interflow}
\eeqa
where $\delta_I$ is a bump 6-form localized at the four-dimensional 
intersection, and $F_1$, $F_2$ are the curvatures of the gauge bundles on 
the D6-brane stacks. Namely there is an inflow of chiral anomaly from the
two sets of D6-branes towars their intersection.

On the other hand, the intersection of D6-branes leads to a chiral 
four-dimensional fermion, localized at the intersection. Such state can be
seen to arise by direct quantization of the sector of open strings stretched 
between the D6-brane stacks, and therefore transforms in the bi-fundamental 
representation of the unitary gauge groups on the D6-branes. The triangle
anomaly associated to that fermion is
\beqa
[\, \ch(F_1) \, \ch(F_2)\, \hat{A}(R)\,]^{(1)}
\eeqa
and is localized at the intersection. Hence it precisely cancels the above 
anomaly inflow (\ref{interflow}).

\medskip

For future comparison, it is interesting to consider the inflow mechanism 
in compact models with intersecting branes. For instance, we consider type
IIA theory compactified on a Calabi-Yau threefold ${\bf X}_3$, with stacks 
of $N_a$ D6-branes wrapped on 3-cycles ${\bf \Pi}_a$. Models of this kind 
with ${\bf X}_3 = \IT^6$ (or orbifolds/orientifolds thereof) as the 
internal space have been considered e.g. in \cite{orbif, 
bgkl,afiru,rest,susy}, and the following inflow picture was described in 
\cite{afiru}.

Localized anomalies due to chiral fermions at the intersections of 
D6-branes are cancelled by the above inflow mechanism. However, since the 
D6-brane worldvolume (transverse to the four-dimensional intersections) is 
compact, global consistency requires that the inflows at different 
intersections add up to zero (inflows to some intersections are 
compensated by outflows from other intersections). This constraint, which 
follows from cancellation of RR tadpoles, implies 
that the overal anomalies from fermions at intersections cancel in the 
ordinary four-dimensional sense. In particular, cubic
non-abelian triangle anomalies automatically vanish, while $U(1)$ -
non-abelian and $U(1)$-gravitational mixed anomalies 
cancel via a Green-Schwarz mechanism \cite{afiru,susy}.

\subsection{Chirality from fluxes}
\label{chiral}

In this section we turn to a novel situation where the inflow mechanism 
plays a key role, but contrary to the previous case does not imply the 
appearance of dynamical chiral fermion degrees of freedom.

\subsubsection{The inflow}

The basic observation stems form the fact that certain combinations of 
NS-NS and RR fluxes couple to RR fields in the same way as D-branes do. In 
particular, type II theories contain couplings
\beqa
\int_{{\bf X}_{10}} \, H_{NS} \,  G_{6-p} \,  C_{p+1}
\label{source}
\eeqa
with $G_{6-p}$ the field strength of the RR field $C_{5-p}$, and $p$ 
even/odd for IIB/IIA theories. Hence, configurations with net flux 
for $H_{NS} G_{6-p}$ are sources for the RR field $C_{p+1}$. The 
coupling is normalized such that one unit of flux carries one unit of 
D$p$-brane charge. A familiar example of such coupling is the type IIB 
interaction $\int H_{NS} H_{RR} C_4^+$, which endows $H_{NS} H_{RR}$ with 
D3-brane charge.

Following the general argument above, combinations of fluxes acting as 
sources for RR fields may induce anomaly inflows on D-branes present in 
the background. For concreteness, let us consider a (massive) type IIA 
configuration with non-trivial value for the 0-form field strength 
$\lambda$, i.e. cosmological constant, and $H_{NS}$ flux. Due to the coupling
\beqa
\int_{{\bf X}_{10}}\, \lambda \, H_{NS} \, C_{7}
\eeqa
the flux acts as a source for the RR field $C_7$, namely
\beqa
dG_2 \, = \, \lambda \, H_{NS}
\eeqa
where $G_2$ is the fields stregnth of $C_1$, the IIA RR 1-form.

Consider a D6-brane in this background. Among its Chern-Simons couplings, 
written as in (\ref{CS2}), its world-volume action contains
\beqa
S_{D6}\, =\, \int_{D6} \, G_2 \, [\, \ch(F)\,  \hat{A}(R)^{1/2}\,]^{(0)}
\eeqa
Which is not gauge invariant, its gauge variation being given by
\beqa
\delta \, S_{D6}\, = \, \int_{D6} dG_2 \,  [\, \ch(F)\, \hat{A}(R)^{1/2}\, 
]^{(1)}  \, = \,
\int_{D6} \,\lambda \, H_{NS}  \, [\, \ch(F)\, 
\hat{A}(R)^{1/2}\, ]^{(1)}
\eeqa

Namely, in the presence of suitable flux, there exists an anomaly inflow 
from the D6-brane volume towards the four-dimensional core of the $H_{NS}$ 
flux background, see fig \ref{fig1}b. In general $H_{NS}$ may not be 
bumpy, so we interpret the 
density of inflow anomaly at a given four-dimensional slice is the 
magnitude of $\lambda H_{NS}$ at such slice.

It is straightforward to cook other configurations leading to inflows 
towards two-, four-, six-dimensional subspaces. Given the structure 
of (\ref{source}) and the world-volume Chern-Simons coupling
\beqa
\int_{Dq}\, G_{8-p}  \,\, Y^{(0)}(F,R)
\eeqa
a D$q$-brane in the presence of longitudinal $H_{NS} G_{6-p}$ flux 
develops anomaly inflow towards $(p+q-8)$-dimensional subpaces. Hence for 
instance, a D7-brane in the presence of longitudinal $H_{NS}$, $H_{RR}$ 
3-form fluxes develops inflow towards 2-dimensional subspaces; a D9-brane 
in the presence of $H_{NS}$ and 1-form field strength fluxes develops 
inflows towards six-dimensional subspaces.

\subsubsection{Absence of dynamical chiral fermions}

The question we would like to address is how this anomaly is cancelled, 
namely what effect takes place at the core of the flux background. 
Notice that the anomaly inflow is `almost' that necessary to cancel the 
anomaly of a fermion localized at the core of the $H_{NS}$ flux. This is 
true for the pure gauge anomalies, since the inflow of anomaly equals 
$[\ch(F)]^{(1)}$, a fermion anomaly. However, the pure/mixed gravitational 
anomaly is not quite right to be cancelled by a dynamical fermion degree 
of freedom, being ${\hat A}^{1/2}$ instead of ${\hat A}$. This is a first 
hint suggesting the anomaly will not be 
cancelled by a dynamical chiral fermion \footnote{Recall that in the case 
of intersecting D-branes the gravitational anomaly is fully reproduced 
thanks to the inflow from both D-branes towards the intersection. In the 
present case, one of the contributions if lacking due to the absence of 
Chern-Simons couplings in the `volume' of the flux.}.

A second piece of evidence in this direction is that a dynamical 
fermion with anomaly cancelling against the inflow should be transforming 
in the fundamental of the D-brane gauge field, and be a singlet under any 
other gauge factor. This contrasts with the familiar situation where 
degrees of freedom associated to D-branes transform in {\em bi}-fundamental 
or adjoint representations. 
This observation excludes the possibility of obtaining the 
desired chirality from a fermion arising by from a non-zero index of the 
Dirac operator of D-brane worldvolume fermions, 
since these transform in adjoint representations, and so would whatever 
fermion zero modes they give rise to.

In fact, since fundamental representations arise from open strings with 
one endpoint on the D-brane, the main problem is where is the other 
endpoint? Despite the lack of a solvable worldsheet CFT in the 
presence of RR fluxes, the worldsheet description is valid and reliable, 
at least for dilute fluxes. Hence an additional endpoint must exist, and 
it is highly unlikely that it somehow `ends' on, or dissipates into, the flux.

In fact this issue can be studied in more detail by considering a 
situation where the RR flux admits a geometric description in M-theory. 
To this purpose, consider one D6-brane along the directions 0123456, in 
the presence of one unit of longitudinal $H_{NS}$ and $F_2$ fluxes, along 
234 and 56, respectively. In the M-theory lift, the former lifts to a 
$G_4$ flux turned on along the directions 234 and the M-theory circle. To 
simplify the rest of the lift, consider the directions 56 to parametrize 
a 2-sphere, pierced by the nonzero $F_2$ flux. The 2-sphere lifts to a 
3-sphere, with the twist in the Hopf fibration reproducing the $F_2$ flux. 
Moreover, to reproduce the D6-brane charge, the M-theory circle should 
degenerate at a point in 789. Hence, the geometry is not simply 
${\bf M}_{8}\times \IS^3$, but rather a fibration of squashed $\IS^3$'s 
over $\IR^3$ (parametrized by 789), times $M_5$ (along 01234). The 
squashing is determined by the property that the circle fiber in $\IS^3$ 
is fibered over $\IR^3$ such that it is a (different) Hopf fibration over 
the angular $\IS^2$ in $\IR^3$. Overall the topology is that of a cone 
over a 5-manifold wich is a circle fibration over $\IS^2\times \IS^2$ with 
Chern classes equal to one. At the tip of the cone only one of the 
$\IS^2$'s goes to zero volume. 

Notice that the geometry is completely smooth (in fact, comparing this
topology with \cite{kw} we learn that the geometry is a resolved
conifold). Since the lift of the configuration is completely smooth,
there is no obvious room to generate a chiral fermion from, say, some 
wrapped M2-brane state, which is the required kind of object to be charged 
in under the gauge field. This is our last piece of evidence for the 
non-existence of dynamical fermions in such situations.

\subsubsection{Origin of the anomaly}

Our proposal to cancel the chiral anomaly inflow is that the flux induces 
an explicitly anomalous term in the D-brane worldvolume, of exactly the right 
form. The key observation is that the D-brane action contains topological 
couplings to the $B_{NS}$ field arising from the Chern-Simons coupling
\beqa
\int_{Dq} \, {\cal G} \, \, [\, e^{(F-B_{NS})} \,
\hat{A}(R)^{1/2}\, ]^{(0)}
\eeqa
For concreteness let us consider the previous situation of a D6-brane in 
the presence of a $\lambda H_{NS}$ background. The D6-brane contains a 
coupling
\beqa
\int_{D6} \,\lambda \, B_{NS} \, [\, \ch(F)\, 
{\hat{A}}^{1/2}\, ]^{(0)}
\eeqa
Notice this term is different from the ones involved in the inflow. 
Moreover, it is explicitly non gauge invariant, its gauge variation being
\beqa
\int_{D6} \, \lambda \, B_{NS}  \, \delta [\, \ch(F)\, 
{\hat{A}}^{1/2}\, ]^{(0)} \, = \,
\int_{D6} \, \lambda \, H_{NS}  \, [\, \ch(F)\, 
{\hat{A}}^{1/2} \, ]^{(1)}   
\eeqa
which is exactly of the form required to cancel the anomaly generated by 
the inflow. 

Notice that this mechanism avoids the problems mentioned above. In 
particular the interaction induced by the presence of the flux provides 
the gauge and gravitational anomalies required to cancel the inflow.

The bottomline is that in the kind of situation we have considered, the 
inflow is cancelled not by a dynamical degree of freedom, but rather by an 
explicitly anomalous interaction induced by the flux.

\medskip

This effect is indeed surprising, in particular in compactified models 
(see section \ref{dsix} below). In fact it allows the construction 
of say four-dimensional models with chiral fermions in an anomalous 
representation of the gauge group. The theory is however consistent due to 
the existence of an explicit Wess-Zumino interaction in the effective 
action, which cancels the residual anomaly. The same applies to pure 
gravitational anomalies in two- and six-dimensional compactifications 
exploiting this mechanism.

Two comments are in order: First, the string construction ensures that 
this Wess-Zumino interaction is of microscopic origin, and is not and 
effective description of a dynamical degree of freedom (i.e. a situation 
along the lines of \cite{farhi}. This situation is indeed surprising and 
to our knowledge not previously discussed in string theory. Second, the 
triangle anomalies cancelled by this mechanism are 
both reducible and irreducible, hence the mechanism is in principle not 
related to a Green-Schwarz anomaly cancellation mechanism (which only 
applies to reducible anomalies).

\section{Applications}
\label{applic}

In this section we turn to two short applications of the effect we have 
discussed.

\subsection{Probe analysis of tadpole vs anomaly cancellation}
\label{probe}

Introduction of an allowed brane probe in a consistent string background 
should result in a consistent world-volume field theory on the probe. In 
\cite{probes} it was proposed that tadpole cancellation conditions 
manifest as anomaly cancellation conditions on suitable brane probes. In 
this section we illustrate than on consistent backgrounds with fluxes 
cancellation of anomalies on brane probes may require the above 
flux-induced Wess-Zumino term in a crucial way.

Consider a prototypical example of compactification with fluxes, for 
instance F-theory compactified on a Calabi-Yau fourfold ${\bf X}_4$, 
elliptically fibered over ${\bf B}_3$. The $C_4^+$ tadpole 
$N_\chi=-\chi({\bf X}_4)/24$ in \cite{svw} can be cancelled by a number 
$N_{D3}$ of D3-branes and/or turning on NS-NS and RR 3-form fluxes with 
$N_{H}=\int_{B_3} H_{NS} H_{RR}$, with $N_\chi + N_{D3} + N_{H}=0$.

Following \cite{probes} one may test tadpole cancellation by introducing a 
D7-brane probe (or a stack of $n$ such D7-probes) wrapped on the base 
${\bf B}_3$, and testing the cancellation of two-dimensional gauge 
anomalies on the non-compact piece of the 
D7-probe worldvolume. The curvature of $B_3$, and the background type IIB 
$(p,q)$ 7-branes induces a gauge anomaly of $N_{\chi}$ units on the 
two-dimensional non-compact piece of the D7-brane worldvolume (see below 
for an explicit example), in conventions where the anomaly of a left-handed 
Majorana-Weyl fermion in the fundamental of $U(n)$ is $+1$. In the absence 
of fluxes, the intersections between $N_{D3}$ D3-branes and the D7-probe lead 
to $N_{D3}$ left-handed Majorana-Weyl fermions in the fundamental, so that 
the total anomaly cancels, since $N_{\chi}+N_{D3}=0$. In the presence of 
$N_{H}$ units of flux, there are not enough fermions form intersections 
between the D3-branes and the D7-probe. Happily, the D7-probe effective 
action also contains an explicit Wess-Zumino term, induced by the presence 
of the flux, via the coupling
\beqa
\int_{D7}\, B_{NS}\, H_{RR}\, [\, \ch(F)\, \hat{A}(R)^{1/2}\,]^{(0)} 
\eeqa
In the effective 2d probe theory this coupling gives rise, to a term 
$N_{H} {\cal L}_{WZ}$, with ${\cal L}_{WZ}$ the Wess-Zumino lagrangian
\footnote{By Wess-Zumino lagrangian here we refer to any non-gauge
invariant interaction whose gauge variation is the appropriate
anomalous term (see e.g. \cite{manes}. It is typically defined by picking 
a one dimension higher manifold ${\bf X}_{D+1}$ with boundary the 
spacetime of interest, and has the typical form ${\cal S}_{WZ}= 
\int_{{\bf X}_{D+1}}\, Y(0)(F,R)$. By default we consider Wess-Zumino 
terms with gauge fields traced in the fundamental representation}. This 
non gauge invariant term cancels the remaining anomaly, and renders the 
probe worldvolume theory consistent.

\medskip

An explicit example is given by taking ${\bf X}_4$ to be ${\bf K3} 
\times {\bf K3}$. 
Equivalently, type IIB theory compactified on ${\bf K3} \times \IP_1$ with 
24 $(p,q)$ 7-branes wrapped on ${\bf K3}$ and located at points in 
$\IP_1$. Since $\chi({\bf X_4})=24\times 24$, there exists a $C_4^+$ 
tadpole of $N_{\chi}=-24$ which must be cancelled by introducing D3-branes 
and/or fluxes with $N_{D3}+N_H=24$. 

Let us recover this condition by analyzing cancellation of gauge 
anomalies on a D7-brane probe wrapped on ${\bf K3} \times \IP^1$. The 
intersections of the D7 probe with the $(p,q)$ 7-branes lead each to a 
chiral 6d fermion propagating on ${\bf K3}$. Due to the non-zero index of 
the Dirac operator on ${\bf K3}$, each such six-dimensional fermion gives 
a chiral right-handed MW two-dimensional fermion, so that we get a total 
two-dimensional anomaly of $-24$ on the D7-brane probe worldvolume. In 
the absence of flux, this anomaly can only be cancelled by dynamical 
fermions arising from intersections between D3-branes and 
the D7-brane probe, hence one requires $N_{D3}=24$. In the presence of 
$N_H$ units of flux, we are left with $-24+N_{D3}$ units of 2d anomaly, 
but the flux-induced Wess-Zumino term contributes an additional amount of 
$N_H$ units of two-dimensional anomaly, hence consistency of the probe 
world-volume theory requires $N_{D3}+N_H=24$, and reproduces the tadpole 
condition.

It is straightforward to consider other compactifications where fluxes 
contribute to tadpole cancellation, and verify the crucial role of the 
flux-induced WZ terms in the physics of D-brane probes. In fact, the 
present work originated from the study of the interplay between tadpole 
cancellation and anomaly cancellation on brane probes in compactifications 
with fluxes.

\subsection{Application to model building}
\label{dsix}

Type II compactifications with RR and NS-NS fluxes are a promising class 
of models whose exploration is still in progress. In this section we 
illustrate that in models leading to chiral 
lower-dimension theories, the flux-induced WZ terms play a crucial role in 
understanding anomaly cancellation from the lower-dimensional viewpoint.

In order to illustrate this point, we consider a particular class of 
models. For concreteness we center on compactifications of type IIA theory 
on Calabi-Yau threefold ${\bf X}_3$ with $N_a$ D6-branes wrapped on 
homology 3-cycles $[\Pi_a]$, and NS-NS flux $H_{NS}$ turned on with total 
homology class $[H_{NS}]$, and in the presence of a cosmological constant 
$\lambda$. Compactifications of this kind in the absence of fluxes have 
been considered in \cite{orbif,bgkl,afiru,rest,susy}. Notice that these 
compactifications are typically non-supersymmetric, but supersymmetric 
configurations could be obtained by introducing O6-planes \cite{susy}. 
Study of supersymmetric models lies beyond the scope of the present 
paper, which centers on more topological aspects.

Extending the analysis in \cite{afiru}, we now derive the RR tadpole 
cancellation conditions. The action for the RR 7-form field is
\beqa
S \, = \, \int_{M_4\times X_3} \, dC_7  \, * dC_7 \, + \, 
\sum_a \, N_a \, \int_{M_4\times \Pi_a} \, C_7 \, + \, \int_{M_4\times 
X_3} \, \lambda \, H_{NS} \, C_7 
\eeqa
The equation of motion is
\beqa
dH_2\, = \, \sum_a \, N_a \, \delta(\Pi_a) \,+ \, \lambda \, H_{NS}
\eeqa
where $H_2$ is the field strength of the RR 1-form, and $\delta(\Pi_a)$ is 
a bump 3-form on ${\bf X}_3$ with support on $\Pi_a$. The equation in 
homology reads
\beqa
\sum_a \, N_a \, [\, \Pi_a \, ] \, + \, \lambda \, [\, H_{NS} \, ] \, = \, 
0
\label{tad}
\eeqa

The low energy four-dimensional theory is generically chiral, with chiral 
fermions arising from D6-brane intersections. The gauge group is $\prod_a 
\, U(N_a)$, and the chiral fermion content is given by
\beqa
\sum_{a<b} \, I_{ab} (\fund_a,\antifund_b)
\eeqa
where $I_{ab}=[\Pi_a]\cdot [\Pi_b]$ is the intersection number, with its 
sign specifying the fermion chirality.

In the presence of fluxes, the above fermion content is explicitly 
anomalous. In particular, the $SU(N_a)^3$ cubic anomalies are given by
$\sum_b I_{ab}$ which need not vanish. Happily we are now familiar with 
the presence of explicit flux-induced four-dimensional Wess-Zumino terms, 
in this case arising from the coupling $\int_{D6}\, B_{NS} Y^{(0)}$. In 
the present case the four-dimensional anomalous interaction has the form
\beqa
S_{WZ} \, = \, \sum_a \,  I_{aH} \, \int_{M_4} \, {\cal L}_{WZ,a}
\eeqa
where $I_{aH}=\lambda \, [\Pi_a]\cdot [H_{NS}]$ and ${\cal L}_{WZ,a}$ is 
the Wess-Zumino lagrangian corresponding to a fermion in the fundamental 
of $U(N_a)$ (including the $U(1)$ factor), and singlet under the remaining 
factors.

The total $SU(N_a)^3$ cubic anomaly is therefore $\sum_b I_{ab}+ I_{aH}$, 
which cancels by using the tadpole condition (\ref{tad}). It is easy to
check that mixed $U(1)$ - non-abelian anomalies cancel by a Green-Schwarz
mechanism, exactly as in \cite{afiru}, while mixed gravitational 
anomalies vanish automatically.

\medskip

It is easy to find applications of the use of fluxes in compact model
building, to avoid the appearance of unwanted gauge groups and matter
contents associated to extra D-brane stacks, typically present in all
realistic D-brane compactifications. For instance, in the realistic models
obtained using intersecting stacks of D6-branes, there exist spectator
sets of D6-branes whose presence is only required by tadpole cancellation,
but which lead to additional gauge factors and matter content.
Substitution of these D6-branes by a $\lambda H_{NS}$ flux in the
appropriate homology class would improve the models (see last reference 
in \cite{rest}). 

A more interesting application of this idea would be to use the
replacement of branes by fluxes to avoid certain chiral, but anomaly-free
combinations of fields present in the model. For instance, the
Standard Model like construction in \cite{susy} contains three standard
quark-lepton generation plus an additional chiral but anomaly-free set of
exotic fields. It would be interesting to get rid of these fields by
replacing some of the D-branes by suitable fluxes.

\section{Brane-flux transitions and the fate of chiral fermions}
\label{braneflux}

The topological couplings (\ref{source}) show that certain combinations of
NS-NS and 
RR fluxes are endowed with charge under RR fields, mimicking the coupling 
of a D$p$-brane. This fact supports the possibility of transitions between 
branes and configurations of suitable fluxes, in a way consistent with 
charge conservation.

Such transitions indeed occur, and are typically non-perturbative, i.e. 
the configurations have to tunnel through some potential barrier, see 
\cite{kpv} for a nice description of one such effect. Moreover, and even 
though it has not been exploited in the literature, we would like to 
emphasize that in some instances a continuous interpolation is also 
possible. To see that, consider the M-theory lift of a IIA process where 
an instanton dissolved into a stack of D6-branes turns into a dynamical 
D2-brane, via a continuous small instanton transition. In the M-theory 
lift, a multi-Taub-NUT space with a 4-form flux proportional to a 
harmonic form in Taub-NUT times an instanton emits the flux as a dynamical 
M2-brane, in a process we may refer to as `small fluxon transition'. By 
performing a 9-11 flip in the above process, it can be described purely 
in terms of fluxes and branes in string theory
\footnote{One may worry that the instanton configuration involves gauge 
non-abelian degrees of freedom, hence involves non-perturbative states 
wrapped on collapsing cycles. However, one may discuss the transition 
involving only abelian instantons, obtained by compactifying additional 
dimensions in a two two-tori and turning on (abelian) magnetic fields on 
them.}.

A similar but non-supersymmetric transition takes place in the 9-11
flipped version of the process where a D4 brane along say 01234 approaches
a D6-brane along 0123456 and is dissolved into a vortex-like magnetic
flux in the D6-brane worldvolume. In the flipped version, a NS5-brane
approaches a Taub-NUT space and is dissolved into a $G_4$ flux of the form
$\omega \wedge F$, where $\omega$ is the harmonic 2-form in Taub-NUT and
$F$ denotes the gauge field in the vortex configuration.

\medskip

Hence brane-flux transitions occur, either by tunneling or in a 
continuous fashion. Given this, it is natural to expect that in the 
presence of additional D-branes (probes) they may induce transitions 
where chiral fermions disappear from  the probe world-volume, and explicit 
flux-induced Wess-Zumino terms arise, in order to maintain the 
world-volume anomaly cancellation. These processes interpolate between 
figures \ref{fig1}a and \ref{fig1}b.

It is straightforward to provide explicit example of the above process, by 
considering transitions in F-theory compactifications on four-folds where 
some D3-branes are turned into fluxes. Introducing D7-brane probes as in 
section \ref{probe} would lead an explicit construction where chiral 
fermions 
disappear from the two-dimensional field theory, leaving behing a 
compensating flux-induced explicit WZ term. One may operate similarly in 
configurations of the kind studied in section \ref{dsix}, or other 
analogous construction.

\medskip

It is interesting to compare the situation with other chirality changing 
phase transitions. Phase transitions in string theory in which the chiral 
content changes by (dis)appearance of fields in a non-anomalous 
representation have been studied in \cite{transi1, transi2,susy}. Since 
the anomaly structure of the theory is unchanged, such transitions do not 
require any compensating WZ terms. It is amusing that our flux-induced 
mechanism allows the occurrence of transitions in which the dynamical 
degrees of freedom with (dis)appear transform in anomalous 
representations.

\section{Global gauge anomalies}
\label{global}

It is easy to device configurations of branes and fluxes which lead to
flux-induced global gauge anomalies on the D-brane world-volume. In
this section we discuss one such example. The construction is closely
related to the question of describing fluxe sources with RR charge 
classified by K-theory but not by cohomology.

The construction is based on the observation that global gauge anomalies
due to dynamical fermions may arise at intersections of D-branes with
D-branes carrying discrete K-theory charge. Consider a system introduced
in \cite{probes}. We take type I compactified on $\IT^2$ with non-BPS
D7-branes transverse to it. Each of the latter carries a $\IZ_2$ charge in 
K-theory \cite{kth} \footnote{The instability of the D7-brane to decay to 
a bundle on the background D9-branes \cite{bergman,oscar} is irrelevant, 
since it will not be present in the T-dual model of interest studied 
below.}. The discrete charge can be detected by introducing probes, given 
by a stack of D5-branes wrapped on $\IT^2$. At the intersection of the 
D5-branes with each non-BPS D7-brane there arises a four-dimensional Weyl 
fermion in the fundamental of the $USp$ group on the D5-brane, hence 
generating a global gauge anomaly. Consistency requires the number of 
D7-branes to be even, corresponding to the cancellation of RR charge in 
full K-theory \cite{probes}.

Let us consider a T-dual version of this construction, namely type IIA on
a $\IT^5$ (with 56789 compactified) with O6$^-$-planes along 013456 and D4
- anti-D4 pairs along 01234 (they may be separated in 789 from the
O6-planes). This system again carries a non-trivial K-theory charge, which
can be detected by introducing D8-brane probes along 012356789. The 
D8-brane gauge group is symplectic and each four-dimensional intersection 
with a D4 (and its image intersection with an anti-D4) leads to a 
four-dimensinoal Weyl fermion in the fundamental, hence to a $\IZ_2$ 
global gauge anomaly.

Our last step is to transform some branes into fluxes using ideas in
section \ref{braneflux}. In fact, considering a D4 - anti-D4 brane well
separated from the O6-plane, there is no topological obstruction to
transforming the D4-brane into a $F_2\, H_{NS}$ flux, and the
anti-D4-brane into its image flux. They carry opposite charge under the RR 
5-form $C_5$, as they should since it is odd under the orientifold action.
However the construction guarantees that the combination of both fluxes
carries a non-trivial $\IZ_2$ K-theory charge. Using straightforward
arguments about cancellation of K-theory charge in the compact space, or
cancellation of global gauge anomalies on the D8-brane
four-dimensional non-compact world-volume, one concludes there exists a
flux-induced $\IZ_2$ global gauge anomaly on the D8-brane worldvolume. 
If the number of pairs turned into fluxes is odd, the overall flux-induced 
global gauge anomaly is cancelled by the dynamical fermions arising from 
the remaining intersections.

Unfortunately the lack of a simple description of world-volume couplings
to the discrete pieces of RR-fields when described in K-theory does not
allow for a more explicit description of this effect. Hopefully the formal
developments in \cite{mw} may help in addressing this issue. It would also
be interesting to find a more general description (or other examples) of
fluxes carrying discrete K-theory charge.

\section{Origin of chirality in string theory}
\label{conclu}

We would like to conclude with a different viewpoint on the relevance of
the mechanism we have described in this paper. One of the basic features
of Particle Physics is chirality, and hence one very interesting question
in String Phenomenology is what String Theory mechanism reproduces
four-dimensional chirality. There are several known mechanism to generate
four-dimensional chiral fermions out of different string/M theory
constructions, some of which have been exploited in building models with
semi-realistic gauge and chiral matter spectrum. Among them, we would like
to mention:

\begin{itemize}
\item a) Compactification of high dimensional gauge interactions with
non-zero index of the Dirac operator (that is, with non-trivial holonomy
in the compactification space, or non-trivial gauge bundle over it). The
prototypical example is heterotic or Horava-Witten compactifications on
Calabi-Yau threefolds \cite{chsw}.

\item b) Type II/I D-branes wrapped on manifolds with non-trivial
tangent/normal bundle and/or non-trivial gauge bundle. An example is
provided by type I compactifications on Calabi-Yau threefolds (e.g.
\cite{sagnotti}).

\item c) D-branes at singularities, the prototypical example being
D3-branes at three-fold orbifold singularities in the transverse space
\cite{dgm}.

\item d) D-branes wrapped on intersecting cycles, the chiral fermions
arising at such intersections \cite{bdl}. The prototypical case is
compactifications with D6-branes wrapped on 3-cycles in the internal
space.

\item e) Isolated $G_2$ holonomy singularities in M-theory, where chiral
fermions arise from M2-branes wrapped on collapsed 2-cycles \cite{aw}.

\item f) Configurations of NS5-branes and D5-branes, known as brane box
and brane diamond models \cite{bbox}.

\item g) D4-branes in the presence of intersections of D6-branes and
NS5-branes \cite{chiral}

\end{itemize}

These different mechanisms give rise to chirality in D-brane world-volume
in a similar way, namely via the appearance of chiral fermions in the
spectrum. In fact it is easy to argue that, despite their superficial
differences, these mechanism are closely related, in particular they are
often related by string dualities \footnote{I am grateful to Tom\'as
Ort\'{\i}n for raising this question.}, hence should have a unified
description in whatever microscopic theory underlies string theory
\footnote{In a sketchy way, models in a) and b) are often related by
heterotic/type I duality; models in c) are often mapped to b) by
T-duality; models in d) are related to b) and c) by mirror symmetry; the
M-theory lift of models in d) belong to class e); models in f), g) are
T-dual to D-branes in certain orbifold and orientifold singularities
\cite{pu} in c).}. This underlying equivalence, previously unnoticed in
the literature, has a very satisfactory  implication. Namely that the
question of what is the `right' mechanism in string theory to reproduce
the chirality of Particle Physics turns into the question of what of the
above is the most efficient description of the unique underlying
mechanism (say, for the value at which the moduli happen to stabilize).
This is reminiscent of the argument by which duality turns the problem of
selecting a string theory as the `right' one into a problem of moduli
stabilization in a particular perturbative limit.

In this respect, the flux-induced anomaly mechanism we have described in
the present paper provides a qualitatively different source of chirality
in string theory. The fact that it is qualitatively different is clear in
that it does not imply a dynamical fermion degree of freedom in the
spectrum, hence no duality can relate it to the above mechanisms. In fact,
it would be interesting to apply duality relations to our D-brane
configurations to discover dual realizations of the mechanism.
Interestingly, we have seen that even though there is no equivalence to
the above mechanisms, the flux-induced anomaly is connected to them by
brane/flux transitions, as discussed in Section \ref{braneflux}. This is
reminiscent of the way different compactifications of a string theory are 
connected by topology-changing transitions. Hence, in a sense it turns the 
problem of what is the origin of chirality into a choice out of two 
topologically different mechanisms, which are neverthelss related by 
physical transition.

\centerline{\bf Acknowledgements}

I thank L.~E.~Ib\'a\~nez for useful discussions and the ITP, Santa
Barbara, for hospitality during completion of this work. I also thank
M.~Gonz\'alez for kind support and encouragement. This work is supported
by the Ministerio de Ciencia y Tecnolog\'{\i}a (Spain) through a Ram\'on y
Cajal contract. This research was also supported in part by the National
Science Foundation under Grant No. PHY99-07949.


\bigskip

\bigskip

\begin{thebibliography}{99}

\bibitem{ps}
J.~Polchinski, A.~Strominger, `New vacua for type II string theory', 
Phys. Lett. B388 (1996) 736, hep-th/9510227.

\bibitem{bb}
K.~Becker, M.~Becker, `M theory on eight manifolds', Nucl. Phys. B477 
(1996) 155, hep-th/9605053.

\bibitem{michelson}
J.~Michelson, `Compactifications of type IIB strings to four-dimensions 
with nontrivial classical potential', Nucl. Phys. B495 
(1997) 127, hep-th/9610151.

\bibitem{gvw}
S.~Gukov, C.~Vafa, E.~Witten, `CFT's from Calabi-Yau four folds',
Nucl. Phys. B584 (2000) 69, Erratum-ibid. B608 (2001) 477, hep-th/9906070.

\bibitem{drs}
K.~Dasgupta, G.~Rajesh, S.~Sethi, `M theory, orientifolds and G-flux', 
JHEP 9908 (1999) 023, hep-th/9908088.

\bibitem{tv}
T.~R.~Taylor, C.~Vafa, `R R flux on Calabi-Yau and partial supersymmetry
breaking', Phys. Lett. B474 (2000) 130, hep-th/9912152.

\bibitem{gss}
B.~R.~Greene, K.~Schalm, G.~Shiu, `Warped compactifications in M and F 
theory', Nucl. Phys. B584 (2000) 480, hep-th/0004103.

\bibitem{cklt}
G.~Curio, A.~Klemm, D.~Lust, S.~Theisen, `On the vacuum structure of type 
II string compactifications on Calabi-Yau spaces with H fluxes',
Nucl. Phys. B609 (2001) 3, hep-th/0012213.

\bibitem{gkp}
S.~B.~Giddings, S.~Kachru, J.~Polchinski, `Hierarchies from fluxes in 
string compactifications', hep-th/0105097.

\bibitem{kst}
S.~Kachru, M.~Schulz, S.~Trivedi, `Moduli stabilization from fluxes in a 
simple iib orientifold', hep-th/0201028.

\bibitem{fp}
A.~R.~Frey, J.~Polchinski, `N=3 warped compactifications', 
hep-th/0201029.

\bibitem{rs}
L.~Randall, R.~Sundrum, `A Large mass hierarchy from a small extra 
dimension', Phys. Rev. Lett. 83 (1999) 3370, hep-ph/9905221.

\bibitem{dielectric}
R.~C.~Myers, `Dielectric branes', JHEP 9912 (1999) 022, hep-th/9910053.

\bibitem{inflow}
M.~B.~Green, J.~A.~Harvey, G.~W.~Moore, `I-brane inflow and anomalous 
couplings on d-branes', Class. Quant. Grav. 14 (1997) 47, hep-th/9605033.

\bibitem{bdl}
M.~Berkooz, M.~R.~Douglas, R.~G.~Leigh, `Branes intersecting at angles',
Nucl. Phys. B480 (1996) 265, hep-th/9606139.

\bibitem{orbif}
R.~Blumenhagen, L.~Gorlich, B.~Kors, `Supersymmetric 4-D orientifolds of 
type IIA with D6-branes at angles', JHEP 0001 (2000) 040, hep-th/9912204; \\
S.~Forste, G.~Honecker, R.~Schreyer, `Supersymmetric Z(N) x Z(M) 
orientifolds in 4-D with D branes at angles', Nucl. Phys. B593 (2001) 
127, hep-th/0008250.

\bibitem{bgkl}
R.~Blumenhagen, L.~Goerlich, B.~Kors, D.~Lust, `Noncommutative 
compactifications of type I strings on tori with magnetic background 
flux', JHEP 0010 (2000) 006, hep-th/0007024.

\bibitem{afiru}
G.~Aldazabal, S.~Franco, L.~E.~Ibanez, R.~Rabadan, A.~M.~Uranga, `D=4 
chiral string compactifications from intersecting branes',
J. Math. Phys. 42 (2001) 3103, hep-th/0011073; `Intersecting brane 
worlds', JHEP 0102 (2001) 047, hep-ph/0011132.

\bibitem{rest}
R.~Blumenhagen, B.~Kors, D.~Lust, `Type I strings with F flux and B flux',
JHEP 0102 (2001) 030, hep-th/0012156; \\
L.~E.~Ibanez, F.~Marchesano, R.~Rabadan, `Getting just the standard model 
at intersecting branes', JHEP 0111 (2001) 002, hep-th/0105155; \\
S.~Forste, G.~Honecker, R.~Schreyer, `Orientifolds with branes at angles',
JHEP 0106 (2001) 004, hep-th/0105208; \\
R.~Blumenhagen, B.~Kors, D.~Lust, T.~Ott, `The standard model from stable 
intersecting brane world orbifolds', Nucl. Phys. B616 (2001) 3, 
hep-th/0107138; \\
D.~Cremades, L.~E.~Ibanez, F.~Marchesano, `SUSY Quivers, Intersecting 
Branes and the Modest Hierarchy Problem', hep-th/0201205.

\bibitem{susy}
M.~Cvetic, G.~Shiu, A.~M.~Uranga, `Chiral four-dimensional N=1  
supersymmetric type 2A orientifolds from intersecting D6 branes',
Nucl. Phys. B615 (2001) 3, hep-th/0107166; `Three family supersymmetric 
standard - like models from intersecting brane worlds', Phys. Rev. Lett. 
87 (2001) 201801, hep-th/0107143.

\bibitem{kw}
I.~R.~Klebanov, E.~Witten,`Superconformal field theory on three-branes at 
a Calabi-Yau singularity', Nucl. Phys. B536 (1998) 199, hep-th/9807080.

\bibitem{farhi}
E.~D'Hoker, E.~Farhi, `Decoupling a fermion whose mass is generated by a 
Yukawa coupling: the general case', Nucl. Phys. B248 (1984) 59.

\bibitem{probes}
A.~M.~Uranga, `D-brane probes, RR tadpole cancellation and K theory 
charge', Nucl. Phys. B598 (2001) 225, hep-th/0011048.

\bibitem{svw}
S.~Sethi, C.~Vafa, E.~Witten, `Constraints on low dimensional string 
compactifications', Nucl. Phys. B480 (1996) 213, hep-th/9606122.

\bibitem{manes}
J.~L.~Ma\~nes, `Differential geometric construction of the gauged 
Wess-Zumino action', Nucl. Phys. B250 (1985) 369.

\bibitem{kpv}
S.~Kachru, J.~Pearson, H.~Verlinde, `Brane/flux annihilation and the
string dual of a nonsupersymmetric field theory', hep-th/0112197.

\bibitem{transi1}
S.~Kachru, E.~Silverstein, `Chirality changing phase transitions in 4-D 
string vacua', Nucl. Phys. B504 (1997) 272, hep-th/9704185; \\
G.~Aldazabal, A.~Font, L.~E.~Ibanez, A.~M.~Uranga, G.~Violero, 
`Nonperturbative heterotic D = 6, D = 4, N=1 orbifold vacua', Nucl. Phys. 
B519 (1998) 239, hep-th/9706158.

\bibitem{transi2}
B.~A.~Ovrut, T.~Pantev, J.~Park, `Small instanton transitions in heterotic 
M theory', JHEP 0005 (2000) 045, hep-th/0001133.

\bibitem{kth}
E.~Witten, `D-branes and K theory', JHEP 9812 (1998) 019, hep-th/9810188.

\bibitem{bergman}
O.~Bergman. `Tachyon condensation in unstable type I D-brane systems',
JHEP 0011 (2000) 015, hep-th/0009252. 

\bibitem{oscar}
O.~Loaiza-Brito, A.~M.~Uranga, `The Fate of the type I nonBPS D7-brane', 
Nucl. Phys. B619 (2001) 211, hep-th/0104173.

\bibitem{mw}
G.~W.~Moore, E.~Witten, `Selfduality, Ramond-Ramond fields, and K theory',
JHEP 0005 (2000) 032, hep-th/9912279; D.~S.~Freed, M.~J.~Hopkins, `On
Ramond-Ramond fields and K theory', JHEP 0005 (2000) 044, hep-th/0002027.

\bibitem{chsw}
P.~Candelas, Gary T. Horowitz, Andrew Strominger, Edward Witten, `Vacuum
configurations for superstrings', Nucl. Phys. B258 (1985) 46.

\bibitem{sagnotti}
C.~Angelantonj, M.~Bianchi, G.~Pradisi, A.~Sagnotti, Ya.~S.~Stanev,
`Chiral asymmetry in four-dimensional open string vacua', Phys. Lett. B385
(1996) 96, hep-th/9606169.

\bibitem{dgm}
M.~R.~Douglas, B.~R.~Greene, D.~R.~Morrison, `Orbifold resolution by
D-branes', Nucl. Phys. B506 (1997) 84, hep-th/9704151.

\bibitem{aw}
M.~Atiyah, E.~Witten, `M theory dynamics on a manifold of G(2) holonomy', 
hep-th/0107177.

\bibitem{bbox} 
A.~Hanany, A.~Zaffaroni, `On the realization of chiral four-dimensional
gauge theories using branes', JHEP 9805 (1998) 001, hep-th/9801134;

M.~Aganagic, A.~Karch, D.~Lust, A.~Miemiec, `Mirror symmetries for brane 
configurations and branes at singularities', Nucl. Phys. B569 (2000) 277, 
hep-th/9903093.

\bibitem{chiral}
A.~Hanany, A.~Zaffaroni, `Chiral symmetry from type IIA branes',
Nucl. Phys. B509 (1998) 145, hep-th/9706047.


\bibitem{pu}
A.~Hanany, A.~M.~Uranga, `Brane boxes and branes on singularities', JHEP
9805 (1998) 013, hep-th/9805139. \\
J.~Park, R.~Rabadan, A.~M.~Uranga, `N=1 type IIA brane configurations,
chirality and T duality', Nucl. Phys. B570 (2000) 3, hep-th/9907074.

\end{thebibliography}
\end{document}